# Step size of the rotary proton motor in single $F_oF_1$-ATP synthase from a thermoalkaliphilic bacterium by DCO-ALEX FRET

Eva Hammann[a], Andrea Zappe[a], Stefanie Keis[b], Stefan Ernst[a,c], Doreen Matthies[d], Thomas Meier[d], Gregory M. Cook[b], Michael Börsch[a,c,*]

[a] 3rd Institute of Physics, University of Stuttgart, Pfaffenwaldring 57, 70550 Stuttgart, Germany
[b] Department of Microbiology and Immunology, Otago School of Medical Sciences, University of Otago, 9054 Dunedin, New Zealand,
[c] Single-Molecule Microscopy Group, Jena University Hospital, Friedrich Schiller University Jena, Nonnenplan 2 - 4, 07743 Jena, Germany
[d] Department of Structural Biology, Max-Planck Institute of Biophysics, Max-von-Laue-Str. 3, 60438 Frankfurt/Main, Germany

**ABSTRACT**

Thermophilic enzymes can operate at higher temperatures but show reduced activities at room temperature. They are in general more stable during preparation and, accordingly, are considered to be more rigid in structure. Crystallization is often easier compared to proteins from bacteria growing at ambient temperatures, especially for membrane proteins. The ATP-producing enzyme $F_oF_1$-ATP synthase from thermoalkaliphilic *Caldalkalibacillus thermarum* strain TA2.A1 is driven by a $F_o$ motor consisting of a ring of 13 *c*-subunits. We applied a single-molecule Förster resonance energy transfer (FRET) approach using duty cycle-optimized alternating laser excitation (DCO-ALEX) to monitor the expected 13-stepped rotary $F_o$ motor at work. New FRET transition histograms were developed to identify the smaller step sizes compared to the 10-stepped $F_o$ motor of the *Escherichia coli* enzyme. Dwell time analysis revealed the temperature and the LDAO dependence of the $F_o$ motor activity on the single molecule level. Back-and-forth stepping of the $F_o$ motor occurs fast indicating a high flexibility in the membrane part of this thermophilic enzyme.

## 1 INTRODUCTION

The membrane-embedded enzyme $F_oF_1$-ATP synthase is ubiquitous in all kind of living cells. It is located in plasma membranes of bacteria, in thylakoid membranes of the chloroplasts in plant cells, and in the inner mitochondrial membranes of mammalian cells. The main purpose is to catalyze the synthesis of adenosine triphosphate (ATP) which is the central energy currency for almost all processes of the living cell. It has been postulated by P. Boyer[1] and later shown experimentally that an internal rotation of subunits is causing a sequential and concerted opening and closing of the three nucleotide binding sites where adenosine diphosphate (ADP) and phosphate ($P_i$) are bound in the presence of $Mg^{2+}$ [2-7]. The electrochemical potential of protons (or $Na^+$ ions in some organisms[8]) across the membrane is utilized as the driving force for the catalysis of the reaction ADP + $P_i$ → ATP according to the chemiosmotic theory proposed by P. Mitchell[9]. Because ATP can be hydrolyzed by reversing the reaction in the same binding pocket, $F_oF_1$-ATP synthases are regulated by different means in the cells to prevent uncontrolled waste of ATP. The mitochondrial enzyme is prevented from ATP hydrolysis by a specific binding protein $IF_1$ which is thought to block rotation mechanically[10]. The chloroplast enzyme is redox-regulated, and the bacterial enzyme is controlled by a conformational change of the rotary ε-subunit[11] that blocks the rotation of the γ-subunit. A model of the thermoalkaliphilic TA2.A1 $F_oF_1$-ATP synthase with subunit composition is shown in Figure 1A.

---

* m.boersch@physik.uni-stuttgart.de; phone (49) 3641 933745; fax (49) 3641 933750; http://www.m-boersch.org

*C. thermarum* strain TA2.A1 is a thermoalkaliphilic bacterium that grows at a temperature of 65°C at pH 9.5 [12]. *C. thermarum* grows on fermentable carbon sources, like sucrose, over a broad pH range (pH 7.5 to 10.0). However, growth on non-fermentable carbon sources such as succinate or malate is not observed until the external pH is > 9.0, suggesting that non-fermentative thermoalkaliphilic growth is specialized to function at high pH values, but not at pH values near neutral[13]. During growth at high pH, a proton-coupled $F_oF_1$-ATP synthase is utilized by strain TA2.A1 to synthesize ATP despite the lack of protons at pH 9.5 and the suboptimal proton-motive force generated due to the large inverted pH gradient[12]. The $F_oF_1$-ATP synthase from strain TA2.A1 ($TA2F_oF_1$) is specifically adapted to function under these conditions and exhibits several features that are unique to alkaliphilic growth namely, latent ATPase activity[14-16], *a*-subunit modifications[17] and a larger oligomeric *c*-ring[18-20]. The $F_o$ motor is driven by the electrochemical potential difference of protons. The rotary motor comprises a ring of 13 *c*-subunits[19]. The thermoalkaliphilic $TA2F_oF_1$ can be genetically modified and expressed in other bacteria for purification, for example in *Escherichia coli*[17]. In contrast to the regular *E. coli* $F_oF_1$-ATP synthase, ATP hydrolysis is suppressed in the $TA2F_oF_1$ at room temperature, but is activated in the presence of the detergent lauryldimethylamine-oxide LDAO[14].

We applied a single-molecule Förster resonance energy transfer (FRET) approach[21-30] using duty cycle-optimized alternating laser excitation (DCO-ALEX)[27, 31-33] to monitor the expected 13-stepped $F_o$ motor at work. Two fluorophores were attached specifically, one to the static *a*-subunit and another to one of the rotating *c*-subunits in $F_o$. New FRET transition histograms were developed to identify the smaller step sizes compared to the 10-stepped $F_o$ motor of the *E. coli* enzyme. Dwell time analysis and Monte Carlo simulations revealed the temperature and the LDAO dependence of the $F_o$ motor on the single molecule level. The elastic properties of the rotor for this thermophilic enzyme will be compared with the *E. coli* enzyme and discussed.

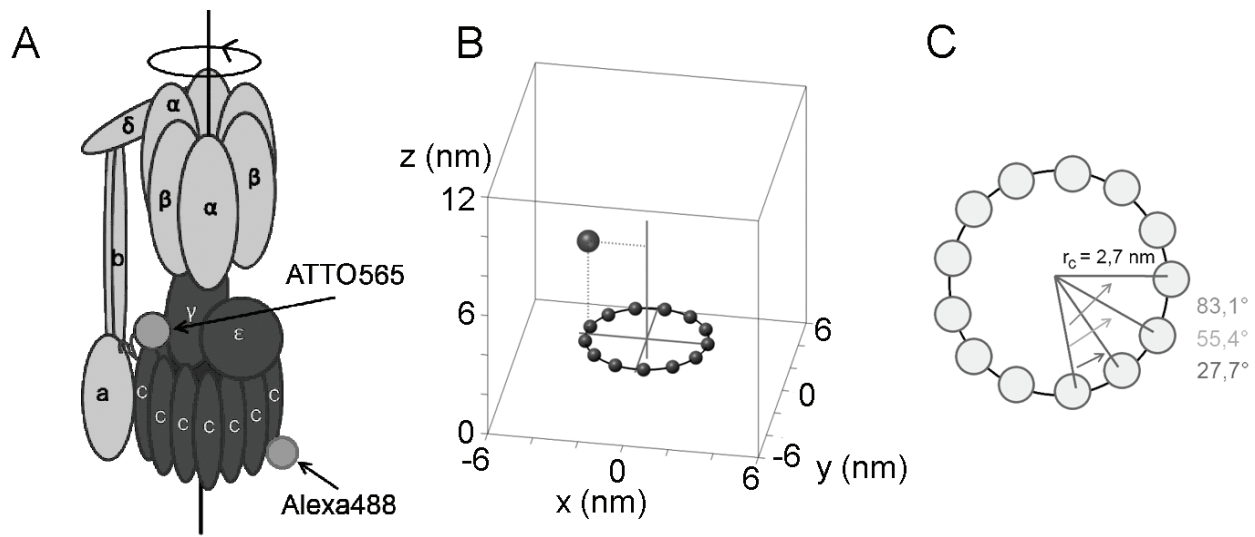

**Figure 1: Model for subunit rotation in $TA2F_oF_1$-ATP synthase.** (A) Subunit composition and rotor / stator assignment. Rotor subunits γ, ε and *c* are shown in dark grey. Alexa-488 is attached to one rotating *c*-subunit, Atto565 is attached to the static *a*-subunit. (B) Structural constraints for *c*-ring rotation measured by single-molecule FRET between the fluorophore on *a* (large dark dot) and one of the 13 *c*-subunits. (C) Defining the step size for single (27.7°) and multiple rotary movements of a *c*-subunit.

# 2 EXPERIMENTAL PROCEDURES

## 2.1 Sample preparation

*Expression and purification of* $TA2F_oF_1$

*Escherichia coli* strain *DK8* [34] harboring the plasmid *psnap* were grown in 2xYT media with the addition of 0.2% glucose and 50 µg/ml ampicillin shaking at 200 rpm at 30°C. At $OD_{600}$ 0.5, the production of $TA2F_oF_1$ was induced by addition of 1 mM isopropyl-β-D-1-thiogalactopyranoside (IPTG). Cells were further incubated for 3 h at 30°C, collected by centrifugation at 5000 x g, immediately frozen in liquid nitrogen and stored at -80°C until purification.
Cells were thawed slowly and washed twice in cell wash buffer (50 mM TRIS-HCl pH 8) at 4°C. The cell pellet was resuspended after centrifugation in French press buffer (20 mM TRIS-HCl (pH 8), 140 mM KCl, 5 mM $MgCl_2$, 5 mM 4-aminobenzoic acid, 10% glycerol (v/v), 5 mM dithiothreitol (DTT) and protease inhibitor (Roche, complete EDTA free)). Cells were disrupted by two passages through a high pressure homogenizer (Avestin). Cell debris was removed by centrifugation (8000 x g, two times 20 min at 4°C). Inverted membrane vesicles were pelleted from the supernatant by ultracentrifugation (434871 x g, 90 min at 4°C). Inverted membrane vesicles were resuspended in cold membrane wash buffer (50 mM TRIS-HCl pH 8, 100 mM KCl 5 mM $MgCl_2$, 5 mM DTT, 10 % (v/v) glycerol, 2% (w/v) saccharose, Roche complete EDTA free) by a brush. Inverted membrane vesicles were pelleted by a second ultracentrifugation step (434871 x g, 90 min at 4°C). Supernatant was carefully removed and the membrane pellet was resuspended in solubilization buffer [20 mM 2-(*N*-morpholino)ethanesulfonic acid (MES) pH 7, 20 mM Tricine, 5 mM $MgCl_2$, 5 mM DTT, 0.05% phenylmethanesulfonylfluoride (PMSF), 1.75 % dodecyl maltoside (DDM)]. 1 g of inverted membrane vesicles were resuspended in 10 ml buffer. $F_oF_1$-ATP synthase was solubilized by stirring 1 h at room temperature. Non-solubilized membranes were removed by ultracentrifugation (311357 x g, 100 min, 4°C). The solubilized $F_oF_1$-ATP synthase was precipitated from the supernatant with 6 M ammonium sulphate (pH 7.4) overnight at 4°C.

*Protein chromatography*

The precipitated protein was collected by centrifugation (8000 x g, 20 min) and was dissolved in binding buffer (20 mM sodium phosphate pH 8, 10 mM imidazole, 5 mM $MgCl_2$, 10 % (v/v) glycerol, 500 mM NaCl). Before applying the protein sample, the 1 ml HisTrap™ FF column (GE Healthcare) was equilibrated with binding buffer. A 5 ml protein sample was loaded on the column and the column was washed with 20 ml of binding buffer. Then the protein was eluted from the column by gradient elution. The imidazole gradient was generated by a mixture of binding buffer and elution buffer (20 mM sodium phosphate pH 8, 500 mM imidazole, 5 mM $MgCl_2$, 10 % (v/v) glycerol, 500 mM NaCl) with raising amounts of elution buffer. The eluted protein was collected in 1 ml samples and precipitated by 6 M ammonium sulphate (pH 7.4). Protein fractions with ATP hydrolysis activity were pooled and dissolved in sample buffer (20 mM sodium phosphate pH 8, 5 mM $MgCl_2$, 10 % (v/v) glycerol, 500 mM NaCl). The protein samples were frozen in liquid nitrogen and stored at -80°C. Protein concentration was determined by amido black staining[35]. The preparation was analyzed by gel electrophoresis using sodium dodecyl sulphate polyacrylamide gel (SDS-PAGE, 12.5%). ATP hydrolysis was measured as described[36].

*labeling and reconstitution into liposomes*

To label $TA2F_oF_1$ specifically with two fluorophores for FRET at the $F_o$ part, we used the Alexa-488 modified ligand BG for the SNAP-tag fused to the non-rotating *a*-subunit according to the instructions of the supplier (New England Biolabs). Briefly, we incubated $TA2F_oF_1$ in 100 µl liposome buffer (20 mM succinic acid pH 8.0, 20 mM Tricine, 60 mM NaCl, 0.6 mM KCl, 0.1 % DDM) at protein concentrations around 1 µM with 10-fold excess of the BG-dye for 1 hour at room temperature. Unbound BG-dye was removed by gel filtration using one or two subsequent Sephadex G-25 spin columns as described[27, 37]. Before labeling of one of 13 cysteines in the *c*-ring TCEP was added to the liposome buffer. We labeled the enzyme with substoichiometric amounts of Atto565-maleimide for 1 h at room temperature. Unbound BG-dye was removed by gel filtration with Sephadex G-50 medium. Specificity of labeling was shown by SDS-PAGE with fluorescence detection of the subunits using a Tyhoon Trio⁺ laser scanner. Afterwards, the FRET-labeled enzymes were reconstituted into preformed liposomes according to published procedures[38-40] to obtain proteoliposomes with not more than a single enzyme per liposome.

## 2.2 Confocal microscope setup with DCO-ALEX

Single-molecule FRET measurements with freely diffusing proteoliposomes (Figure 2A) were accomplished by a custom-built confocal microscope based on an Olympus IX71 stage. Our microscope is a modular system with a variety of lasers and up to 4 single photon counting avalanche photodiodes (APDs; SPCM-AQR-14, Perkin Elmer)[30, 32, 33, 41-44]. The actual configuration is shown in Figure 2B. The FRET donor fluorophore Alexa-488 was excited with a ps-pulsed laser at 488 nm (PicoTA490, Picoquant, Berlin). The laser was triggered for a duty cycle-optimized sequence by a programmable arbitrary waveform generator (AWG 2041, Tektronix) resulting in three pulses within a 96 ns microtime. To prove the existence of the second FRET fluorophore Atto565 at the same enzyme, we switched on a continuous-wave laser at 561 nm (Jive, Cobolt) by an acousto-optical modulator (AOM, Figure 2C) in interleaved or alternating laser configuration (DCO-ALEX)[31, 32, 45-47].

A water immersion objective (UplanSApo 60x, 1.2 N.A.) collected the fluorescence of single proteoliposomes, which was split by a dichroic mirror to separate the FRET donor and acceptor signals. Photons were recorded by two TCSCP PC cards (SPC152, Becker&Hickl, Berlin) and analyzed by our custom-made software "Burst_Analyzer" written by N. Zarrabi (University of Stuttgart). Sorting photons according to the laser pulse sequence was used to generate the three intensity time trajectories with 1 ms binning: i.e. FRET donor, FRET acceptor and directly excited FRET acceptor intensity for threshold-based identification of the photon bursts in the time trajectories (Figure 2D). A temperature-controlled sample chamber was built for single-molecule FRET measurements at temperatures up to 41°C.

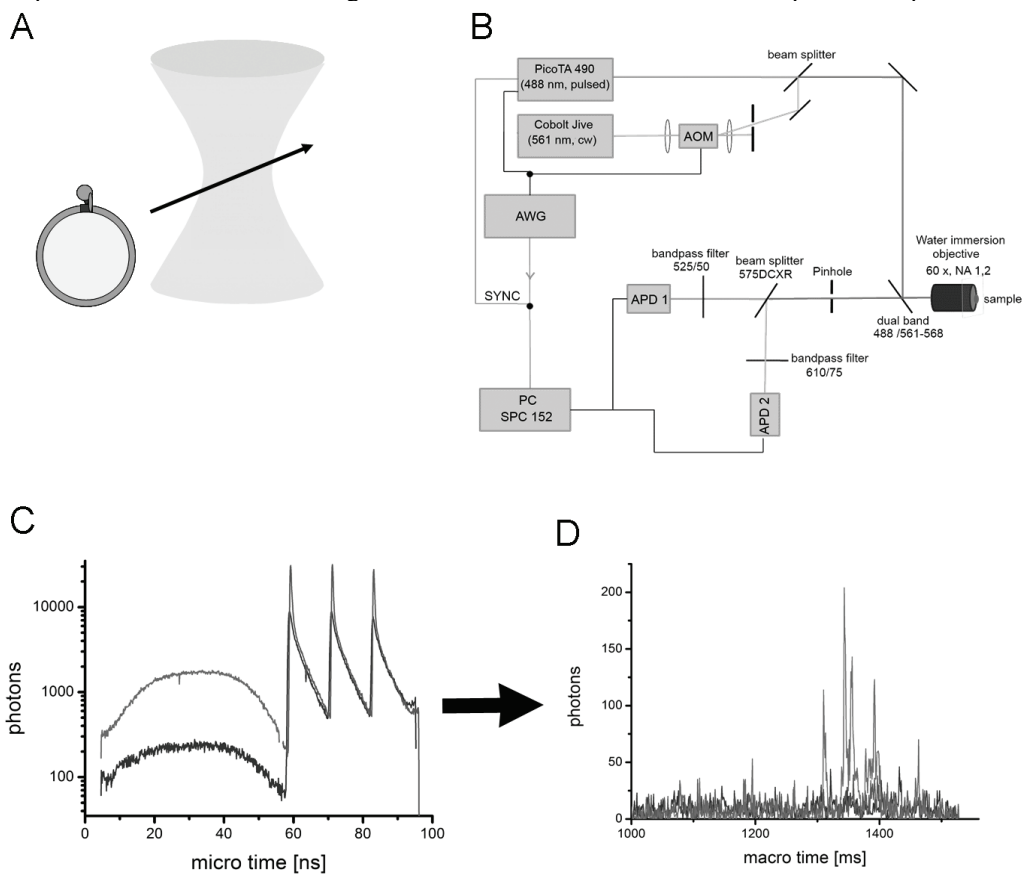

**Figure 2: Monitoring subunit rotation in single** $TA2F_oF_1$. (**A**) The confocal detection approach. A proteoliposome containing a single $F_oF_1$-ATP synthase is freely diffusing through the confocal volume of the lasers drawn as gray cone. (**B**) The microscope set up. Lasers and counting electronics are synchronized by an arbitrary waveform generatorm, AWG. (**C**) Duty cycle-optimized alternating laser excitation scheme. (**D**) Example of a photon burst of a single $TA2F_oF_1$ with three intensity trajectories.

# 3 RESULTS

We constructed a double mutant of TA2.A1 $F_oF_1$-ATP synthase by fusing a SNAP-tag to the C-terminus of the *a*-subunit in $F_o$, and by introducing a cysteine in the N-terminus of the 13 *c*-subunits for single-molecule FRET measurements of *c*-ring rotation. The thermoalkaliphilic enzyme was expressed in *Escherichia coli* lacking $F_oF_1$-ATP synthase and purified using Ni-NTA columns for binding the $His_6$-modified β-subunits of the $F_1$ sector of the holoenzyme. Labeling of the SNAP-tag with either Alexa-488-BG or Atto565-BG was possible, but our labeling efficiencies were low (less than 10%). In contrast, substoichiometric labeling of one of the 13 cysteines was easily possible. Depending on the fluorophore on the SNAP-tag, we used the complementary FRET dye to yield double-labeled enzymes with Alexa-488 and Atto565. The proteins were reconstituted into liposomes. ATP hydrolysis measurements at different temperatures, in absence and presence of LDAO, showed that the double mutant was functionally equivalent to the wild type enzyme (see below, Fig. 6A).

To monitor ATP hydrolysis-driven subunit rotation in single FRET-labeled $TA2F_oF_1$ at temperatures above 21°C, we built a custom-designed heating chamber on top of the scanning stage of our microscope. A machined copper block with springs to hold a cover glass was insulated by a PVC housing with cover plate. Heating or cooling of the chamber was achieved by two Peltier elements with external cooling fins. In the PVC lid and within the copper plate, two PT100 sensors were placed for temperature measurements in the air as well as in the copper block. Calibration curves were recorded by comparing the temperatures set in the copper block and in the air, and the temperature obtained in the water droplet on the cover glass. After 20 minutes pre-heating, a temperature accuracy of ± 2° C was maintained in the buffer. Additional microscope objective heating was not available.

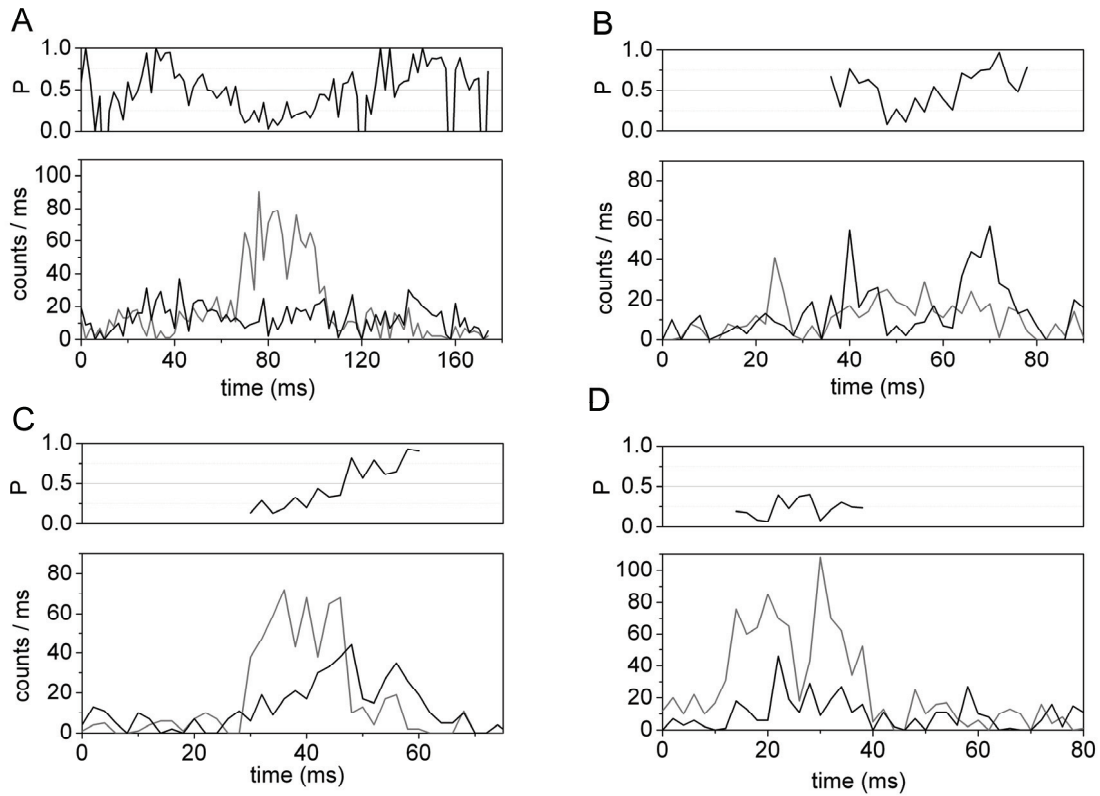

**Figure 3: Photon bursts of single FRET-labeled TA2.A1 $F_oF_1$-ATP synthase during ATP hydrolysis.** (**A**) - (**D**) Gray lines in lower panel correspond to FRET donor intensities (Alexa-488 attached to the *c*-ring), black lines are the FRET acceptor intensities (Atto565 bound to the SNAP-tag on the *a*-subunit). The proximity factor P trace is shown in the corresponding upper panels. (**A**), (**C**) Enzymes with sequential small FRET efficiency changes. (**B**), (**D**) Enzymes with alternating FRET efficiency changes.

Single-molecule FRET measurements of *c*-ring rotation in TA2F$_o$F$_1$ were accomplished in liposome buffer (see Materials and Methods) at pH 8.0 in the presence of 2.5 mM MgCl$_2$ and 1 mM ATP between 21°C and 41°C. Freely diffusing proteoliposomes caused photon bursts (see four examples in Figure 3) during the transit time through the detection volume. Using two alternating lasers enabled us to find those TA2F$_o$F$_1$ in the time trajectories, which contained both fluorophores on the F$_o$ sector, i.e. one Alexa-488 at the *c*-ring and Atto565 at the *a*-SNAP-tag. Manual inspection of each of the these photon bursts allowed us to identify different FRET levels in individual TA2F$_o$F$_1$. FRET efficiencies were calculated from corrected fluorescence intensities (after background subtraction, spectral cross-talk correction, detection efficiency corrections of the set up to yield a correction factor γ=1.2) as the so-called proximity factor P=I$_A$/(I$_D$+I$_A$) shown on Fig. 3. The FRET level histogram of all assigned levels was very broad and covered the complete range from less the 10% up to nearly 100% FRET efficiency (histogram not shown).

Afterwards, the corresponding FRET distances were calculated for each FRET level using the Förster formula[48]. To identify the step size of *c*-ring rotation we computed the FRET transition density plot (Figure 4A). Here, the FRET dis-tance 1 of the first level is plotted against the next FRET level 2. Most of the FRET transitions were found near the diagonal indicating small step sizes for increasing as well as decreasing distances. This was expected for a symmetrical *c*-ring rotation as shown previously for *c*-ring rotation in *E. coli* $F_oF_1$-ATP synthase[27]. To identify the step size from FRET transition density plots, we used the geometrical model shown in Figures 1B and 1C and calculated the distance changes for either single steps (27.7 deg), double steps (55.4 deg) and triple steps (83.1 deg) of the fluorophore in the *c*-ring with respect to the fluorophore at the *a*-subunit. The resulting ellipses were plotted over FRET transitions in Fig. 4A. The inner, smallest ellipse was related to the 27.7 deg stepping, and the outer ellipse to the 83.1 deg stepping. The FRET transition distribution appeared to be dominated by single 27.7 deg steps. However, a significant fraction of data points could also be attributed to double steps (55.4 deg) of the *c*-ring, i.e. single steps were not resolved, caused by our manual assignment of FRET level and the given time resolution of at least 2 ms to identify a distinct FRET level.

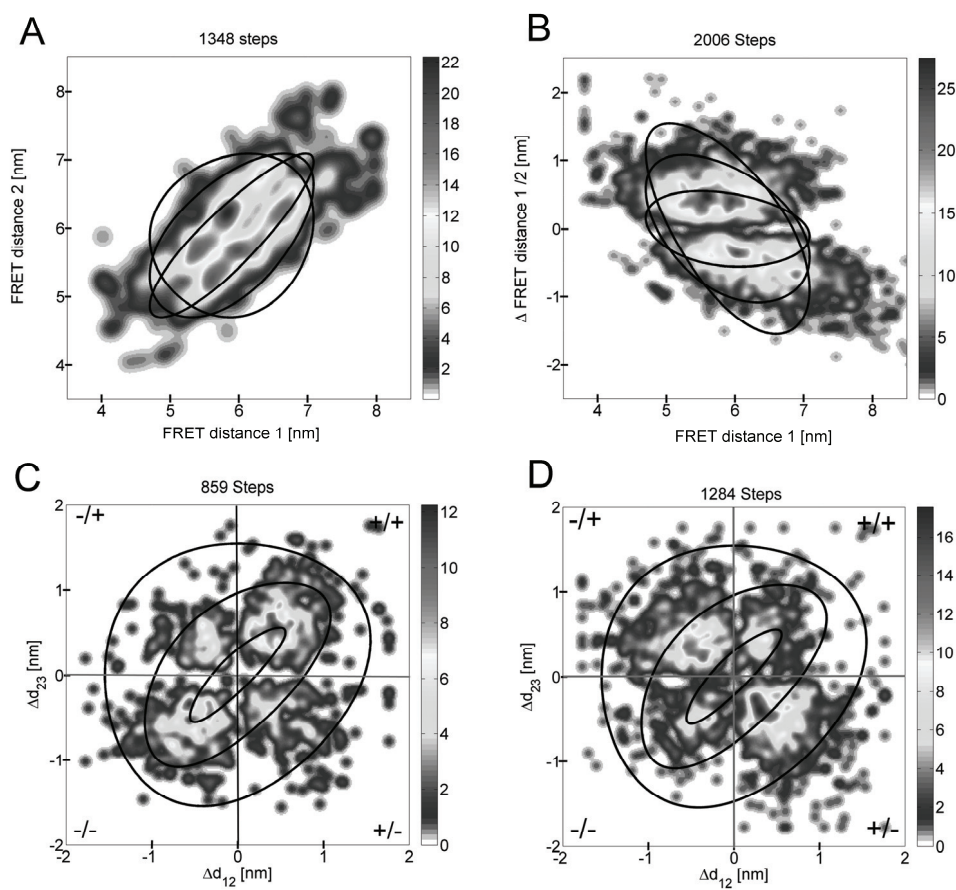

**Figure 4. FRET transitions density plots for single FRET-labeled TA2F$_o$F$_1$ during ATP hydrolysis** showing at least 3 distinct FRET levels. (**A**) FRET transitions between FRET distance 1 and the subsequent FRET distance 2. The black ellipses correspond to expected FRET transitions in a step-by-step motion in either 27.7° [inner ellipse], 55.4° [intermediate ellipse] or 83.1° [outer ellipse] (**B**) New FRET transition diagram by plotting the starting FRET distance 1 versus the distance change Δ$_{12}$ in nm. The ellipses show the expected changes for either 27.5° [inner ellipse], 55.4° [intermediate ellipse] or 83.1° [outer ellipse]. (**C**) Plotting the FRET distance change Δ$_{12}$ versus the subsequent FRET distance change Δ$_{23}$ in photon bursts with three and more FRET levels. The first three FRET level showed sequential unidirectional changes. (**D**) Plotting the FRET distance change Δ$_{12}$ versus the subsequent FRET distance change Δ$_{23}$ in photon bursts with three and more FRET levels. The first three FRET level showed alternating changes.

Therefore we searched for more meaningful FRET transition plots to discriminate the different small distance changes associated with the step size. For each FRET level, we plotted the subsequent FRET distance change in a two dimensional histogram (Fig. 4B). Here, a rotary movement of the *c*-ring by 27.7 deg corresponded to the horizontal inner ellipse, and the 83.1 deg stepping was associated with the larger, inclined ellipse. In addition, we separated all photon bursts with FRET transitions into two groups: either showing sequential FRET transitions in the first three FRET levels with a photon burst (or unidirectional rotation, respectively). The second group contained photon bursts showing alternating back-and-forth movements. In Fig. 4B, the photon bursts with alternating FRET levels were analyzed. Obviously, most FRET changes were related to single 27.7 deg back-and-forth oscillations of the *c*-ring. Very similar distributions were found for the sequential rotations. However, as discussed for the FRET transition density plot for sequential rotational steps (Fig. 4A), a clear discrimination of one *versus* multiple steps was not doubtless.

At last we plotted FRET distance change $\Delta_{12}$ between FRET level 1 and 2 *versus* the subsequent FRET distance change $\Delta_{23}$, i.e. between FRET level 2 and 3. Figure 4C shows the FRET level change distribution for the sequential FRET changes, and Figure 4D for the alternating FRET changes. To analyze these FRET distance change distributions, one has to keep in mind that small FRET level changes of less than 0.5 nm could not be assigned in the single-molecule FRET trajectories. Therefore, data points near 0 nm were missing in the FRET distance change transitions. For the sequential FRET distance changes, the majority of transitions was found for either for positive, increasing FRET distances changes followed by also positive FRET distance changes (upper right quadrant of Fig. 4C), or for negative FRET distances followed by negative FRET distances changes (lower left quadrant of Fig. 4C). The FRET distance changes were small and about ± 0.5 nm corresponding to the inner ellipse for single 27.7 deg stepping of the *c*-ring. However, as the sorting of the photon bursts to "sequential changes" (in contrast to "alternating") was based on the first three FRET level within a multi-level burst, a significant number of transitions were found for apparently "alternating" stepping, that is, in the lower right and upper left quadrants of Fig. 4C.

For the enzymes sorted into "alternating" FRET transition sequences shown in Fig. 4D we found the majority of FRET distance changes in the upper left and lower right quadrants, as expected for back-and-forth stepping or an oscillating *c*-ring. The majority of FRET distance changes appeared at very small distances of less than ± 0.5 nm. As seen in Fig. 4D, a significant number of FRET transitions deviated from a simple back-and-forth fluctuation within the single photon burst. In comparison with Fig. 4C, the apparent sequential FRET distance changes seemed to result more obviously from single 27.7 deg stepping of the *c*-ring and not from multiple fast stepping in a row.

All three types of FRET transition plots of the experimental single-molecule FRET data measured with $TA2F_oF_1$ during ATP hydrolysis indicated at preferential step size of the *c*-ring in the rotary movement of single *c*-subunit. To unravel the limitations of the FRET approach, we computed the results of hypothetical different step sizes of the ring of 13 *c*-subunit by Monte Carlo simulations (Figure 5). We assumed a radius of Alexa-488 rotation in the *c*-ring of 2.7 nm, a distance of 2.6 nm for Atto565 from the axis of rotation, a height between the planes for the two fluorophores of 4.7 nm, a Förster radius of 5.8 nm and a low mean photon count rate for both FRET donor an acceptor of 70 counts / ms. The two dimensional distributions for either single steps (27.7 deg), double steps (55.4 deg). triple steps (83.1 deg) or quadruple steps (110.8 deg) comprised characteristic patterns for each step size. For example, single steps relate to the smallest ellipse in the FRET distance change plot (Fig. 5C), but quadruple steps for the largest ellipse (Fig. 5L). FRET distance changes up to 2 nm were only seen for quadruple steps (Fig. 5K), but not for smaller step sizes.

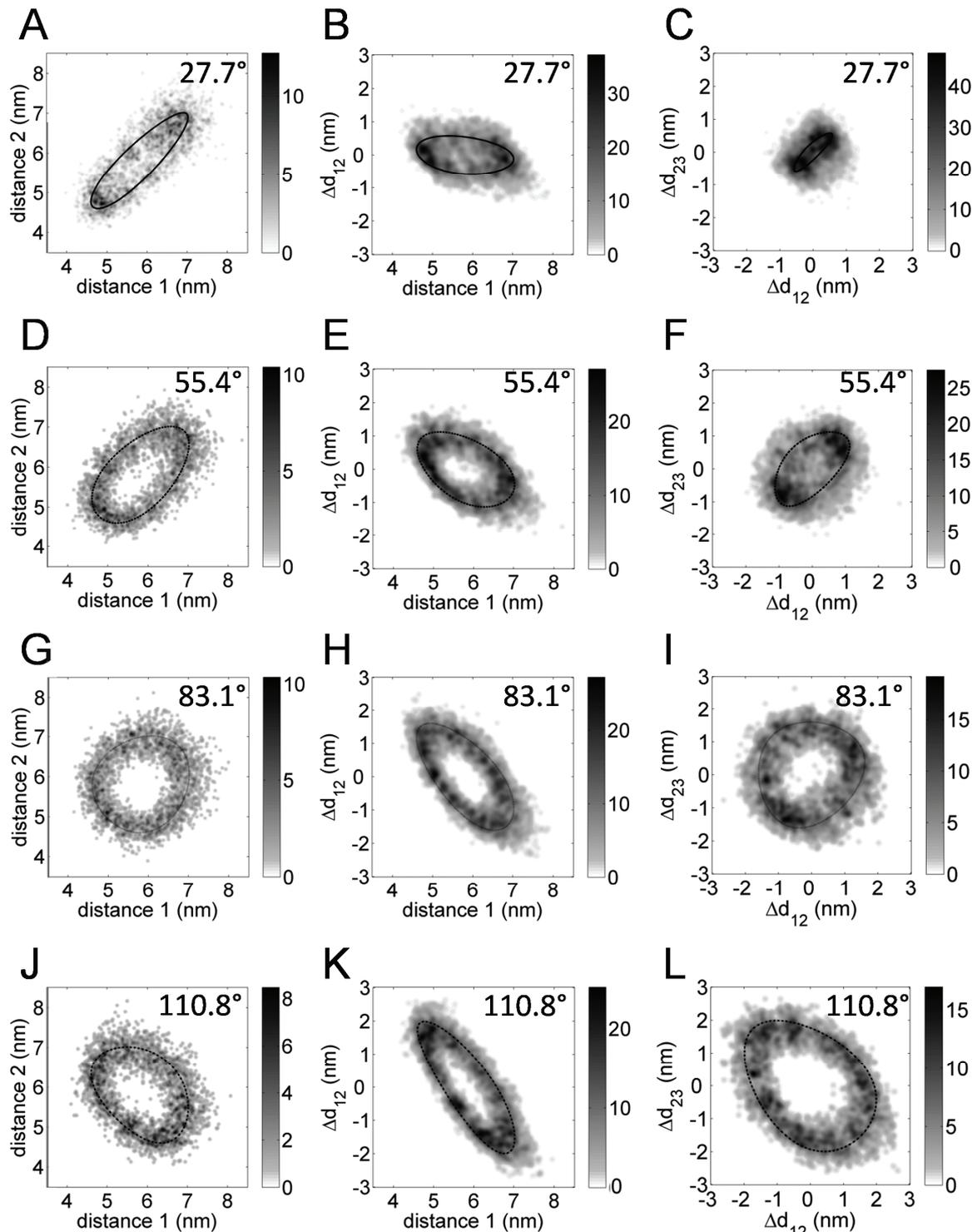

**Figure 5. Monte Carlo simulations of FRET transitions for different step sizes**. (**A**), (**D**), (**G**), (**J**) FRET transition density plots showing pairs of FRET distances. For the MC simulations, given low photon count rates of 70 counts/ms resulted in blurred transitions. (**B**), (**E**), (**H**), (**K**) FRET distance changes for a given FRET distance. (**C**), (**F**), (**I**), (**L**) Pairs of FRET distance changes for a triple-set of consecutive FRET levels in sequential unidirectional order.

The single-molecule FRET measurements of *c*-ring rotation in single reconstituted $TA2F_oF_1$ were repeated at different temperatures and in the presence of 0.1% to 0.5% lauryldimethylamine-oxide LDAO. In biochemical assays, increasing temperature and addition of the detergent LDAO resulted in faster ATP hydrolysis rates of the FRET-labeled enzymes (Figure 6A). At all temperatures as well as in the presence of LDAO we observed fluctuating FRET efficiencies within single photon bursts. To correlate the FRET changes with *c*-ring rotation, we summarized all dwell times in histograms and fitted the distribution with monoexponential decays. Four conditions are shown in Figures 6B - E. The mean dwell time decreased from 4.9 ms at 21°C without LDAO to 3.7 ms at 31°C in the presence of 0.1% LDAO, and further to 2.2 ms at 41°C in the presence of 0.1% LDAO. At least comparable trends were observed for biochemical turnover assays and the dwell time analysis. However, ATP hydrolysis was stimulated much more in the biochemical assays than found in the single-molecule FRET level dwell times.

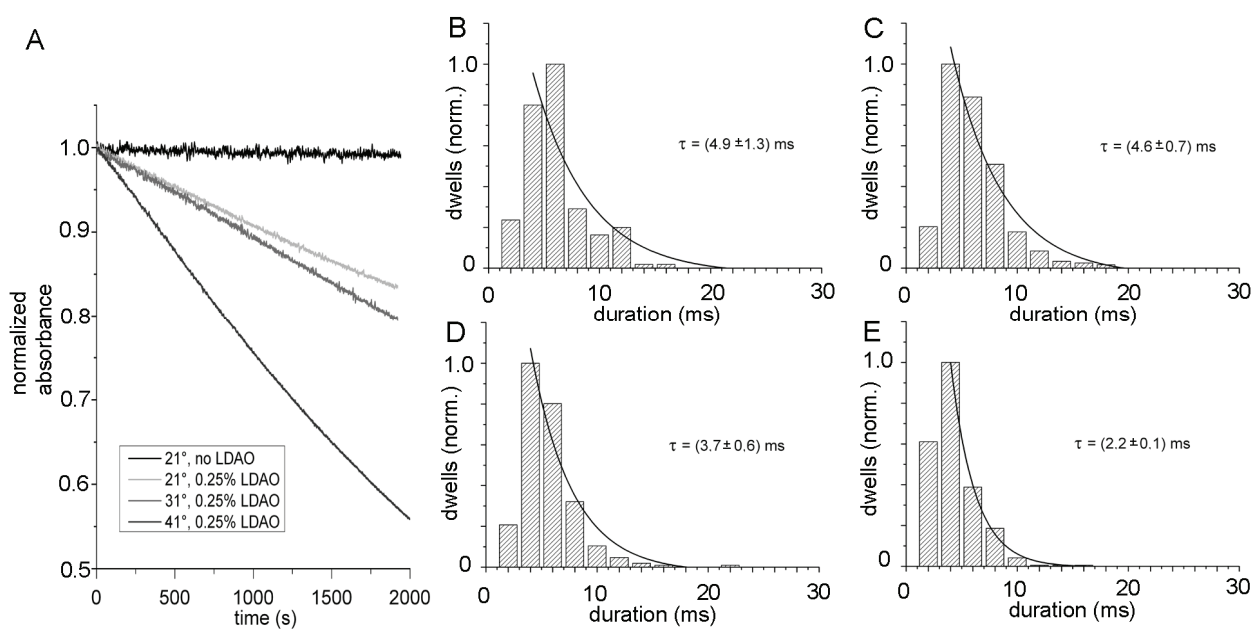

**Figure 6. ATP hydrolysis rates and dwell times. (A)** Temperature and LDAO-dependent ATP hydrolysis rates of reconstituted $TA2F_oF_1$ in absence or presence of LDAO. **(B)-(E)** Normalized dwell time distribution from single-molecule FRET measurements for *c*-ring rotation; in the absence **(B)** of LDAO or in the presence of 0.1 % LDAO **(C, D, E)** at 21°C **(B, C)**, 31°C **(D)** and 41°C **(E)**. Dwell times τ were fitted with monoexponential decays.

Finally, we examined possible pitfalls in our single-molecule FRET analysis. Especially we searched for photophysical artifacts. In the first experiments, manual inspection of photon bursts of Alexa-488-labeled SNAP-tag on the *a*-subunit of $TA2F_oF_1$ indicated a low photon count rate per ms, in contrast to the unbound Alexa-488 fluorophore in buffer solution. Burst-integrated single-molecule fluorescence lifetime measurements[49] were used and the calculated lifetimes of Alexa-488 on the SNAP-tag of $TA2F_oF_1$ were correlated with the mean brightness of the photon burst (Figure 7A). It appeared that the lifetimes were distributed and ranged from 2.5 ns to 4 ns in the absence of a FRET acceptor. Short lifetimes were associated with low mean intensities. Expected lifetimes between 3.5 ns and 4 ns were rarely found, and solely in bright photon bursts. It seemed that the local protein environment with a variety of tryptophans near the binding site for Alexa-488 in the SNAP-tag caused some fluorescence quenching. Therefore, we changed the position of the FRET donor Alexa-488 to a cysteine on the *c*-ring, and used an Atto565-BG ligand for labeling the SNAP-tag instead. For the Atto565 binding to the SNAP-tag we did not observe significant loss of brightness, and, apparently, the labeling efficiency was higher and more reproducible.

Accurate FRET distance measurements require a knowledge of the actual transition dipole orientations between FRET donor and FRET acceptor dye[50]. To probe the relative mobility for each of the two fluorophores, we measured the fluorescence anisotropies of the two dyes attached to reconstituted single $TA2F_oF_1$ independently. We compared the distributions of single photon burst anisotropies with the distributions of the unbound dyes (Figure 7B, C). In addition, anisotropies of single free rhodamine 110 fluorophores (data not shown) and of diluted solutions of erythrosine B were measured as references and used to calibrate our modified confocal microscope set up. Picosecond-pulsed excitation at 488 nm was applied for Alexa-488, rhodamine 110 and erythrosine anisotropies, and cw excitation at 561 nm for Atto565 and erythrosine anisotropies. For Alexa-488 binding at the cysteine in the *c*-subunit of TA2$F_oF_1$ resulted in a high mean anisotropy r ~ 0.23. A mean single-molecule anisotropy around r ~ 0.25 was found for Atto565 binding to the SNAP-tag on the *a*-subunit (Fig. 7B). Both anisotropies were too high in order to assume a free dye mobility and an orientational averaging on the nanosecond time scale.

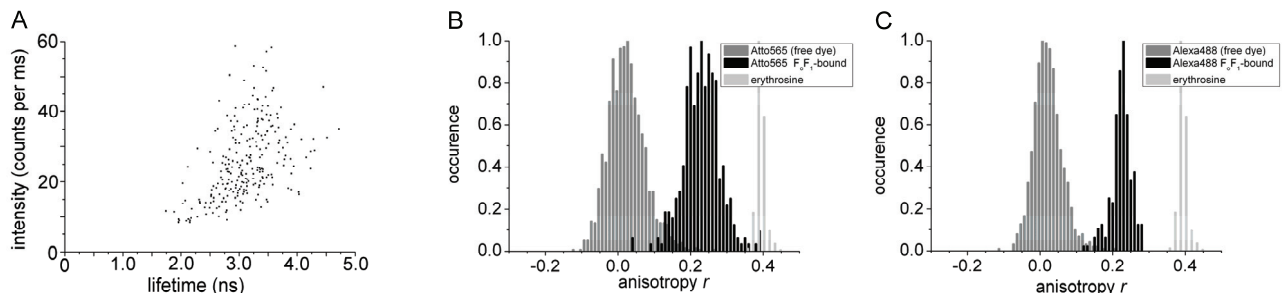

**Figure 7. Fluorescence lifetime distribution of Alexa-488 bound to the SNAP-tag on TA2$F_oF_1$ and single-molecule fluorescence anisotropy distributions.** **(A)** Intensity dependence of Alexa-488 fluorescence lifetimes bound to the SNAP-tag on TA2$F_oF_1$. Burst-integrated fluorescence lifetimes (BIFL)[49] were fitted by a maximum likelihood estimator approach based on a single monoexponential lifetime decay. **(B)** Fluorescence anisotropy distributions of single Atto565 bound to the SNAP-tag on TA2$F_oF_1$, single freely diffusing Atto565 and the reference dye erythrosine in solution. **(C)** Fluorescence anisotropy distributions of single Alexa-488 bound to the SNAP-tag on TA2$F_oF_1$, single freely diffusing Alexa-488 and the anisotropy reference dye erythrosine in solution. **(B)** Linearly polarized cw excitation at 561 nm was used. **(C)** Picosecond-pulsed excitation at 488 nm was used.

## 4 DISCUSSION

$F_oF_1$-ATP synthases are the ubiquitous molecular machines[51] that produce the biochemical energy currency ATP using the electrochemical potential of $H^+$ (or $Na^+$ in some organisms) as the driving force. These enzymes operate at different optimal temperatures, for example, at 37°C in *E. coli* or at 65°C in the thermoalkaliphilic *Caldalkalibacillus thermarum* strain TA2.A1. The ring of *c*-subunits in the $F_o$ sector is considered to act as an "electromotor"[52, 53] transducing mechanical rotation to the 3-stepped motor in $F_1$. The *c*-ring sizes vary from 8 to 15 subunits in different organisms[54-57]. Crystallizations and X-ray structure determinations of a variety of *c*-rings have revealed the diameters, and different lengths and internal distances of the transmembrane helices were found. Especially, the *c*13-ring of TA2$F_o$ contains elongated N- and C-terminal parts of the helices[19, 58].

Here we showed that rotary movements of the ring of 13 *c*-subunits the thermoalkaliphilic TA2$F_oF_1$-ATP synthase can be monitored in real time. Using single reconstituted enzymes labeled with two fluorophores both on the static *a*-subunit and on one rotating *c*-subunit enabled Förster resonance energy transfer measurements. We unraveled sequential and alternating stepping motions with a step size predominantly related to the movement of a single *c*-subunit, that is, by 27.7 deg. The step size did not change at higher temperatures (31°C and 41°C) and in the presence of the detergent LDAO, but the dwell time were shortened at higher temperatures and with LDAO. Whereas the rotation mode in a single step of 27.7 deg is consistent with other single-molecule *c*-rotation measurements[27, 59], the large number of alternating back-and-forth stepping has to be discussed in terms of the thermophilic nature of the *c*13-ring of TA2$F_oF_1$.

More generally, we wanted to address the question how a thermophilic rotor ring differs from a bacterial *c*-ring operating at 37°C. It could be that thermophilic enzymes are stiffer and less flexible to work at higher temperatures[60]. NMR methods are suitable to identify the flexible protein domains and can therefore be used to compare regular and thermophilic proteins. For example, the thermophilic and the mesophilic *E. coli* adenylate kinase have been investigated by NMR[61]. The two enzymes exhibited comparable rates for the closing of the nucleotide binding site induced by ATP, with the thermophilic enzyme being even faster. However, product release was rate limiting and much slower in the thermophilic enzyme. Thus the ATP turnover of the thermophilic enzyme was significantly slower at room temperature than for the *E. coli* enzyme. The faster *c*-ring movement of the TA2F$_o$F$_1$-ATP synthase compared to *E. coli* (~ 4 ms versus ~ 9 ms for a single *c*-subunit step) is therefore not necessarily in contradiction to the very slow ATP hydrolysis rates of the TA2.A1 enzyme at ambient temperatures.

Nevertheless, the limitations of our single-molecule FRET approach should be recalled. The photon count rates were lower than in previous single-molecule FRET experiments using cysteines for labeling[7, 30, 43]. The SNAP-tag used for labeling the *a*-subunit exhibited photophysical quenching effects on Alexa-488 and resulted in high fluorescence anisotropy values similar to a EGFP fusion to the *E. coli a*-subunit[26, 27]. Alternatives could be the use of the Halo-tag, or new ways to produce the F$_o$ sector by combining the individually pre-labeled subunits *a* and *c* in appropriate stoichiometry. The high anisotropies found for the labeled *c*-subunit could be avoided by more hydrophilic fluorophores with longer linkers. Other ideas for improving the FRET-based rotation measurements with single F$_o$F$_1$-ATP synthases have been published recently[62, 63].

The main problem is still the short observation time of the freely diffusing proteoliposome. Expansion of the detection volume will increase the background signal. One very promising method is to confine the diffusion and apply forces to hold the vesicle in given place. A E Cohen (Harvard) and W E Moerner (Stanford) have developed an "Anti-Brownian electrokinetic trap" that uses electrophoretic forces on a vesicle or protein or single fluorophore to trap it in solution[64-69]. They use a confocal microscope with a rotating laser focus in the fastest version of the ABELtrap. Measuring photon count rates of single labeled target in a position-dependent manner allows to generate a fast feedback for voltages applied by four electrodes. Thus the labeled protein is pushed back and the Brownian motion can be completely suppressed. Observation times of up to 10 seconds for single molecules were achieved. This ABELtrap could be used in combination with a single-molecule FRET detection of subunit rotation in F$_o$F$_1$-ATP synthases resulting in several full rotations of a single enzyme and significantly improved statistics to separate photophysical artifacts and to unravel the details of the mechanochemical properties and principles of this remarkable molecular machine.

**Acknowledgements**

This work was in part supported by the DFG grants BO 1891/10-1 and BO 1891/8-1 to M.B., and Marsden Fund, Royal Society to G.M.C. The authors want to thank Prof. Dr. J. Wrachtrup for supporting the Diploma Thesis of E.H. (3$^{rd}$ Institute of Physics, University of Stuttgart). We thank von Gegerfeldt Photonics for the loan of the Cobolt Jive laser.